\documentclass[%
 reprint,
 amsmath,amssymb,
 aps,
prb,
floatfix,
]{revtex4-2}

\usepackage{graphicx}% Include figure files
\usepackage{dcolumn}% Align table columns on decimal point
\usepackage{bm}% bold math
\usepackage{color}  
\usepackage{float}  

\begin{document}

\preprint{v:1.0~~~~~9.12.22}

\title{Spin wave spectra of single crystal CoPS$_3$}% Force line breaks with \\

\author{A. R. Wildes}
 \email{wildes@ill.fr}
\affiliation{Institut Laue-Langevin, 71 avenue des Martyrs CS 20156, 38042 Grenoble Cedex 9, France}

\author{B. F{\aa}k}
\affiliation{Institut Laue-Langevin, 71 avenue des Martyrs CS 20156, 38042 Grenoble Cedex 9, France}

\author{U. B. Hansen}
\affiliation{Institut Laue-Langevin, 71 avenue des Martyrs CS 20156, 38042 Grenoble Cedex 9, France}

\author{M. Enderle}
\affiliation{Institut Laue-Langevin, 71 avenue des Martyrs CS 20156, 38042 Grenoble Cedex 9, France}

\author{J. R. Stewart}
\affiliation{ISIS Pulsed Neutron and Muon Source, STFC Rutherford Appleton Laboratory, Harwell Campus, Didcot, OX11 0QX, UK}

%\author{M. D. Le}
%\affiliation{ISIS Pulsed Neutron and Muon Source, STFC Rutherford Appleton Laboratory, Harwell Campus, Didcot, OX11 0QX, UK}

%\author{R. A. Ewings}
%\affiliation{ISIS Pulsed Neutron and Muon Source, STFC Rutherford Appleton Laboratory, Harwell Campus, Didcot, OX11 0QX, UK}

\author{L. Testa}
\affiliation{Ecole Polytechnique F{\'e}d{\'e}rale de Lausanne, SB ICMP LQM, CH-1015 Lausanne, Switzerland}

\author{H. M. R{\o}nnow}
\affiliation{Ecole Polytechnique F{\'e}d{\'e}rale de Lausanne, SB ICMP LQM, CH-1015 Lausanne, Switzerland}

\author{C. Kim}
\affiliation{Center for Quantum Materials, Seoul National University, Seoul 08826, Korea}
\affiliation{Center for Correlated Electron Systems, Institute for Basic Science, Seoul 08826, Korea}
\affiliation{Department of Physics and Astronomy, Seoul National University, Seoul 08826, Korea}

\author{Je-Geun Park}
\affiliation{Center for Quantum Materials, Seoul National University, Seoul 08826, Korea}
\affiliation{Center for Correlated Electron Systems, Institute for Basic Science, Seoul 08826, Korea}
\affiliation{Department of Physics and Astronomy, Seoul National University, Seoul 08826, Korea}

\date{\today}% It is always \today, today,
             %  but any date may be explicitly specified

\begin{abstract}
The spin waves in single crystals of the layered van der Waals antiferromagnet CoPS$_3$ have been measured using inelastic neutron scattering.  The data show four distinct spin wave branches with large ($\gtrsim 14$ meV) energy gaps at the Brillouin zone center indicating significant anisotropy.  The data were modelled using linear spin wave theory derived from a Heisenberg Hamiltonian.  Exchange interactions up to the third nearest-neighbour in the layered planes were required to fit the data with ferromagnetic $J_1 = -1.37$ meV between first neighbours, antiferromagnetic $J_3 = 3.0$ meV between third neighbours, and a very small $J_2 = 0.09$ meV between second neighbours.  A biaxial single-ion anisotropy was required, with a collinear term $D^x = -0.77$ meV for the axis parallel to the aligned moment direction and a coplanar term $D^z=6.07$ meV for an axis approximately normal to the layered crystal planes.    
\end{abstract}

%\keywords{Suggested keywords}%Use showkeys class option if keyword
                              %display desired
\maketitle

%\tableofcontents

% ***************************
% ****** Introduction ******
% ***************************
\section{Introduction}
Magnetic van der Waals compounds have attracted considerable attention as magnetic analogues to graphene \cite{Park, Li}. These compounds offer possibilities to delaminate down to monolayer thickness and can be intercalated with other atoms and molecules. Of particular interest in this context is the family of (antiferromagnetic) transition-metal compounds \emph{TM}PS$_3$, with \emph{TM}= Mn, Fe, Co, or Ni \cite{Brec, Grasso, Susner, Zhu}. In these compounds, the \emph{TM}$^{2+}$ ions form a honeycomb lattice in the crystallographic $\left(a,b\right)$ planes and have an octahedral coordination with six surrounding sulfur atoms. The composite forms layers that are stacked along the $c$ axis, weakly bound by van der Waals forces,  with the resulting structure corresponding to the monoclinic $C2/m$ space group \cite{Ouvrard85}.

\begin{figure}
  \includegraphics[width=3.5in]{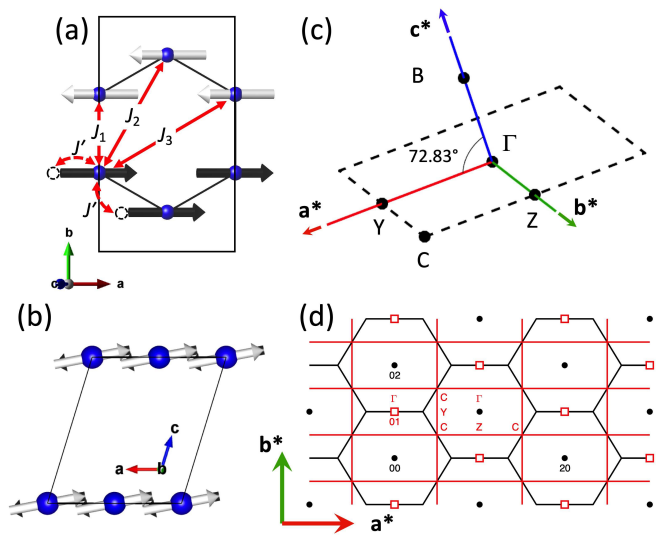}
  \caption{\label{fig:MSBZ} The magnetic structure of CoPS$_3$ \cite{Wildes17}, viewed along the (a) $c^{\star}$ axis, and (b) $b$ axis respectively. The dashed circles in (a) indicate the positions of cobalt atoms in the next $\left(a,b\right)$ plane along the $c$ axis.  The magnetic exchanges between first, second and third nearest intraplanar neighbours are labelled $J_1$, $J_2$ and $J_3$ respectively, and $J^{\prime}$ indicates the interplanar exchange.    The VESTA software package was used to create the figure \cite{VESTA}.  (c) the Brillouin zone for the magnetic structure of CoPS$_3$.  Selected high-symmetry points are indicated, corresponding to the magnetic space group $P_C2_1/m$ (\#11.57) as defined in the Bilbao Crystallographic Library \cite{Bilbao1,Bilbao2,Bilbao3}. (d) an $\left(a^\star, b^{\star}\right)$ plane for CoPS$_3$.  Brillouin zone boundaries are shown for the crystallographic and magnetic lattices as black and red lines respectively.  Nuclear Brillouin zone centres are shown as black filled circles.  Strong magnetic Bragg peaks appear at the red open squares.  Selected \emph{hk} Miller indices and high-symmetry points are indicated.}	
\end{figure}

The quasi-two-dimensionality of the crystal structure extends to the magnetic properties where, even in bulk, the compounds have been investigated as possible model magnetic systems.  CoPS$_3$ deserves particular attention as it is a candidate for possible exotic magnetic states.  Theory models for $S = 3/2$ on a honeycomb lattice, which is the high-spin state for Co$^{2+}$,  predict a possible valence-bond state \cite{Affleck}.  Orbital contributions, also present in Co$^{2+}$, may perturb such a state, however the subsequent lifting of the degeneracy in the $d^7$ orbitals due to spin-orbit and crystal electric field effects combined with lattice distortions may lead to an effective $\mathcal{J}_\textrm{eff} = 1/2$ angular momentum state with associated Kitaev-like physics \cite{Liu, Sano}.  The compound orders antiferromagnetically below a relatively high N{\'e}el temperature of $T_N \sim 120$ K, forming the ``zig-zag" antiferromagnetic  structure shown in figure \ref{fig:MSBZ} \cite{Ouvrard82, Wildes17}.  The structure is similar to other Kitaev-Heisenberg candidates \cite{Trebst} like Na$_2$IrO$_3$ \cite{Choi12}, $\alpha$-RuCl$_3$ \cite{JSears,Ritter}, and the cobalt-containing  compounds Na$_3$Co$_2$SbO$_6$ and Na$_3$Co$_2$TeO$_6$ \cite{CKim_21a}.  Long-ranged magnetic order, which cannot occur in two-dimensional Heisenberg systems according to the Mermin-Wagner theorem \cite{Mermin}, is stabilised by a planar anisotropy that is apparent in the paramagnetic susceptibility \cite{Wildes17}.  The anisotropy is likely associated with orbital contributions but these are relatively small, estimated to be $L \approx 0.32$ based on the effective moment determined from paramagnetic susceptibility ($\mu_\textrm{eff} = 4.55~\mu_\textrm{B}$ versus 3.87 $\mu_\textrm{B}$ for $g = 2$ and spin-only $S = 3/2$) and the size of the ordered moment ($gS = 3.4~\mu_\textrm{B}$ verses 3 for $g = 2$ and $S = 3/2$).

% Figure: Magnetic structure and Brillouin Zone
% end Figure: Magnetic structure and Brillouin Zone

However, neutron inelastic scattering experiments on powdered samples provide the clearest evidence that exotic magnetic states are not found in CoPS$_3$ \cite{CKim, CKim_21b}.  The data indicate that CoPS$_3$ has well-defined spin waves with spectral weight between 15 to 35 meV.  The energy gap is indicative of the relatively large anisotropy.  Spectral weight due to a possible crystal field level was observed at 70 meV \cite{CKim_21b}, but nothing similar appeared at lower energies.  The energy scales for the spin wave spectrum are much larger than those observed for the other Kitaev-Heisenberg candidate compounds, and the scattering could be modelled using linear spin wave theory with a larger spin $S = 3/2$.  The best model required biaxial anisotropy terms, resulting in the spin wave dispersion having four distinct and non-degenerate branches from the four moments per magnetic unit cell.  It was assumed that there was no exchange between the \emph{ab} planes, and that the magnetism was purely two-dimensional.

Powder-averaged data lose the directional information associated with excitation propagation vectors, potentially introducing ambiguity into the modelling.  Measurements on single crystals are necessary to verify that the modelling is correct. This article describes neutron inelastic scattering experiments on co-aligned single crystals of CoPS$_3$ to elucidate the spin dynamics.  The results confirm the conclusions that CoPS$_3$ has the larger $S=3/2$ with clear spin waves, and the data are modelled with linear spin-wave theory.  The results are discussed in the context of the expected electronic configuration of the Co$^{2+}$ in its crystal-symmetry environment.

% ***************************
% ****** Experiment ******
% ***************************
\section{Experiment and results \label{Sec:Exp}}
\subsection{Sample preparation}
Single crystal samples of CoPS$_3$ were grown by direct combination of the pure elements in an evacuated, sealed quartz ampoule using the protocol described in reference \cite{Wildes17}.  The largest crystals, with approximate dimensions $\sim 4 \times 2 \times 0.5$ mm$^3$, were identified and their crystal quality was verified using x-ray Laue diffraction.  The TM-PS$_3$ compounds are well-known to suffer from twinning due to a 120$^\circ$ degree rotation of the \emph{ab} planes around the $c^{\star}$ axis \cite{Murayama}.  All the crystals were determined to be better than 75\% monodomain using neutron diffraction on the IN3 spectrometer at the ILL, France \cite{Wildes_IN3_July19}.

IN3 was also used to co-align the crystals.  Each crystal was individually wrapped in aluminium foil and then attached to a thin aluminium plate using a small amount of GE varnish.  The glue was never in direct contact with the crystals.  The normal to the plate was thus coaxial with the $c^\star$ axis, with the crystals aligned to share the same $b^\star$ axis for their majority domains in the plane of the plate.

A co-aligned sample with four crystals glued to one plate was used for the neutron three-axis spectrometer measurements.  A fifth crystal on a second plate was added for measurements using time-of-flight spectrometry.  The two plates were co-aligned by mounting them on a common alumunium ``toaster rack'' support.

% Three-axis spectrometry
% IN8
\subsection{Unpolarised three-axis spectrometry}
Preliminary neutron measurements were performed using the IN8 thermal neutron three-axis spectrometer at the ILL, France \cite{Wildes_IN8_July19}.  The sample was aligned to measure the scattering in the $\left(hk0\right)$ plane.  IN8 was configured with a bent silicon $111$ monochromator and a horizontally focusing pyrolytic graphite $002$ analyser.  Both monochromator and analyser used vertical focusing.  The final wavenumber was fixed at $k_f = 2.662$ {\AA}$^{-1}$, and a graphite filter was placed before the analyser to filter contamination from higher-order $\lambda/n$ contributions.  Temperature control down to 1.5 K was achieved using a liquid helium cryostat.

\begin{figure}
  \includegraphics[width=3.5in]{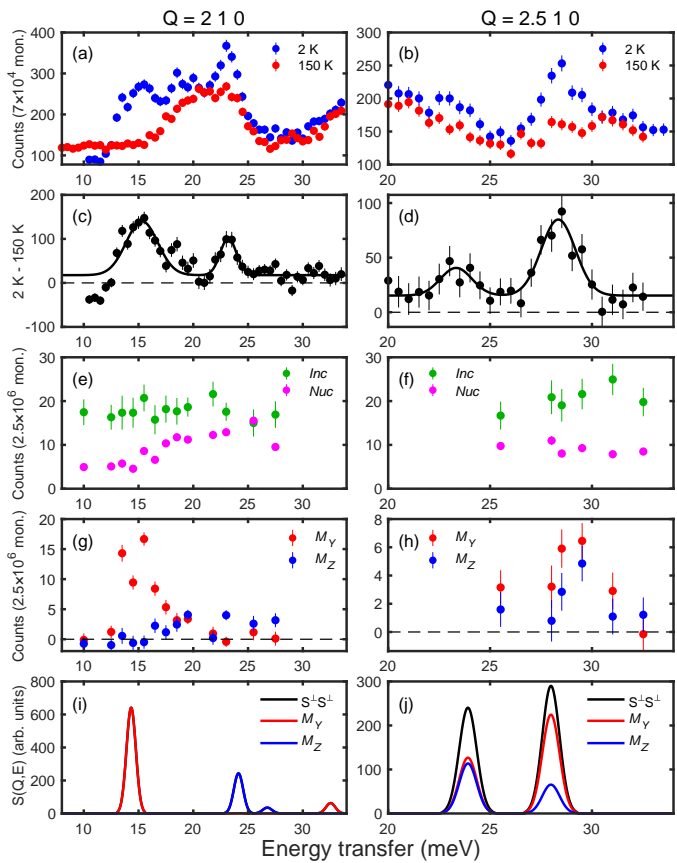}
  \caption{\label{fig:TAS} Neutron three-axis spectroscopy data and simulation for the magnetic excitations in CoPS$_3$ at 210 (left column) and $\frac{5}{2}$10 (right column), corresponding to $\Gamma$ and Y points respectively. 
(a) and (b) show measurements from IN8, each with data at 2 K and 150 K.  The data have been divided by the temperature factor.  
The differences between the scattering at the two temperatures are shown in (c) and (d), and the peak positions have been determined by fitting Gaussian functions.
(e) - (h) show the separated contributions to the scattering determined from IN20 neutron polarisation analysis data using equations \ref{eq:Pol}.  The  nuclear coherent and isotopic incoherent, $Nuc$, and nuclear spin incoherent, $Inc$, contributions are shown in (c) and (d), and the directional components of the magnetic fluctuations, $M_Y$ and $M_Z$, are shown in (g) and (h). 
SpinW calculations of the modelled spin waves are shown in (i) and (j).  The axis definitions for $y$ and $z$ are those used for equation \ref{eq:Hamiltonian_Wildes}, and $S^{\perp}$ refers to the components of the spin that are perpendicular to the neutron scattering vector, $\bf{Q}$.}	
\end{figure}

The measurements confirmed the presence of the energy gap and the band width of the spin wave spectral weight observed in the measurements from powdered samples \cite{CKim}.  However, the scattering appeared to have significant contributions from phonons.  Figures \ref{fig:TAS}(a) and (b) show the scattering measured at 210 and $\frac{5}{2}10$, corresponding respectively to $\Gamma$ and Y points of the magnetic space group $P_C2_1/m$ (\#11.57) shown in figure \ref{fig:MSBZ}(c), in both the magnetically-ordered state at 2 K and well above $T_N$ at 150 K.  The data have been divided by the temperature factor, $1/\left(1-\textrm{exp}\left(-E/k_BT\right)\right)$.  Features are visible at both temperatures.  The high-temperature data underlie the low-temperature data, indicating boson-like features whose energies are relatively temperature-independent, as expected for phononic contributions. 

% Figure: TAS (IN8 & IN20) figure
% end Figure: TAS (IN8 & IN20) figure

The high-temperature data, once corrected for the temperature factor, could then be taken as a phonon background and were subtracted from the low-temperature data to determine the spin-wave contributions.  The temperature subtractions are shown in figures \ref{fig:TAS}(c) and (d).  Two clear peaks at $\sim 15$ meV and $\sim 23$ meV are visible at the $\Gamma$ point.  A suggestion of a third peak at 18.5 meV is likely due to a spurious signal as the elastic scattering momentum triangle with the nominal incident wavenumber at this position closes near to the $2\overline{1}0$ magnetic Bragg peak.  Two peaks are also seen at the Y point, at $\sim 23$ meV and $\sim 28$ meV.  No clear evidence for spin waves was observed below 14 meV or above 30 meV. 

% IN20
\subsection{Polarised three-axis spectrometry}
Neutron polarisation analysis is capable of separating the magnetic scattering from other contributions, for example from phonons, and is furthermore capable of determining the directional components of the magnetic fluctuations.  Consequently, the polarised neutron IN20 spectrometer at the ILL, France was used to clarify the measurements performed on IN8.  

The same sample orientation as IN8 was used for the measurements.  Heusler Cu$_2$MnAl $111$ crystals were used for both the polarizing monochromator and analyser.  The final wavenumber was again fixed at 2.662 {\AA}$^{-1}$ with a graphite filter before the analyser to suppress higher-order contamination.  The neutron polarization direction at the sample position was set using current-carrying coils, and a neutron spin flipper before the analyser was used to measure different polarization states.  A liquid helium cryostat was again used for temperature control.

A total of 6 polarization states were measured at each $\left(\bf{Q},\omega\right)$ point on IN20.  Measurements were performed with the flipper on and off with the polarization axis at the sample in one of three orthogonal directions: parallel to $\bf{Q}$ (designated $X$),  in the scattering plane perpendicular to $\bf{Q}$ ($Y$), and normal to the scattering plane ($Z$).  The measurements with the flipper off and on are designated $NSF_P$ and $SF_P$ respectively for each of the polarization axes $P = X, Y, Z$.  Linear combinations of these measurements were used to separate the different contributions to the total scattering \cite{Moon}, with:

\begin{equation}
	\begin{split}
		&Nuc = \left(2\sum_{P}^{X,Y,Z}NSF_P - \sum_{P}^{X,Y,Z}SF_P\right)/6 \\
		&Inc = 3\left(SF_Y + SF_Z - SF_X\right)/2 \\
		&M_Y = \left(NSF_Y - NSF_X + SF_X - SF_Y\right)/2 \\
		&M_Z = \left(NSF_Z - NSF_X + SF_X - SF_Z\right)/2,
	\end{split}
	\label{eq:Pol}
\end{equation}
where $Nuc$ is the nuclear-coherent and isotope-incoherent scattering, $Inc$ is the nuclear-spin-incoherent scattering, $M_Y$ and $M_Z$ are the scattering due to the magnetic components along $Y$ and $Z$ respectively.  The isotope-incoherent scattering from CoPS$_3$ and the aluminium mount is negligible \cite{Sears} and is ignored in subsequent discussion.

Figures \ref{fig:TAS}(e) and (f) show the separated nuclear-coherent and nuclear-spin-incoherent contributions at the 210 and $\frac{5}{2}10$ positions respectively.  Both are large.  The nuclear-spin-incoherent contribution comes partly from the cobalt, whose incoherent cross-section of 4.8 barns/atom \cite{Sears} is purely due to nuclear spin, and also from the hydrogen in the GE varnish used to glue the crystals.  The large nuclear-coherent scattering verifies the conclusion that the IN8 data suffered from phonon contamination.  The large phonon density at similar energies to the spin waves may have important ramifications regarding magnetostriction properties of CoPS$_3$.

Figure \ref{fig:TAS}(g) shows the purely magnetic scattering at the $\Gamma$ point, and may be directly compared to figure \ref{fig:TAS}(c).   The $\sim$15 meV peak is clearly seen and is shown to be solely due to fluctuations in the $\left(hk0\right)$ plane.  The  $\sim 23$ meV peak and the, probably spurious, peak at $\sim 18.5$ meV are not clearly seen, but the data show fluctuations normal to the $\left(hk0\right)$ plane at these energies with the intensity above 22 meV being exclusively due to these fluctuations.  

Similar measurements at $\frac{5}{2}10$ are shown in figure \ref{fig:TAS}(h), which may be compared to the unpolarised data in figure \ref{fig:TAS}(d).  The measurements concentrated on higher energies.  The peak at $\sim 28$ meV is seen and shown to have contributions from magnetic fluctuations both within and normal to the $hk0$ plane, with the former fluctuations giving a larger intensity.

% Time-of-flight spectrometry
\subsection{Time-of-flight spectrometry}
Broad surveys of reciprocal space were measured using the direct geometry time-of-flight spectrometer MAPS at the ISIS facility, UK \cite{MAPS}.  The instrument was configured with an incident energy of $E_i = 70$ meV and the `S' chopper at a frequency of 250 Hz, giving a resolution of $\Delta E \sim 3$ meV and $\sim 2.2$ meV at energy transfers of 15 meV and 35 meV respectively.  Temperature control to 5 K was achieved with a closed circuit cryorefrigerator.  

The samples were mounted with the $b^{\star}$ axis oriented vertically and were initially aligned with $c^{\star}$ parallel to the incident beam.   A series of measurements were performed as a function of the rotation about the vertical axis, sweeping $\pm 60^{\circ}$ about the initial alignment with $1^{\circ}$ steps.  The data were then combined using MANTID \cite{MANTID} and cuts and slices were then extracted and analysed using the HORACE software \cite{Horace}.

The data showed clear dispersive features within the energy range determined from measurements on powdered samples \cite{CKim}.  Their intensities decreased with increasing $\bf{Q}$, as expected for neutron magnetic scattering due to the form factor, and hence they were taken to be due to spin waves.  The data had less contamination from non-magnetic signals than the three-axis measurements, largely as MAPS was able to access smaller $\bf{Q}$ where the phonon signal is small.

\begin{figure}
  \includegraphics[width=2in]{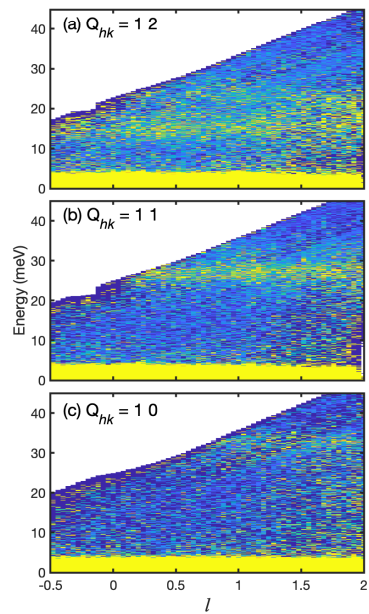}
  \caption{\label{fig:Ldisp} Neutron inelastic scattering data from MAPS for slices parallel to the $c^{\star}$ axis.  Equivalent reciprocal lattice points were shown to be qualitatively the same and the data have been combined for better statistics. The trajectories have fixed $\{hk\}$ of  (a) $\{12\}$, (b) $\{11\}$ and (c)$\{10\}$, thus incorporating Brillioun zone centres.  The slice widths are $\pm 0.1$ in both $h$ and $k$.}	
\end{figure}

Slices parallel to the $c^{\star}$ axis were extracted to determine whether there was any dispersion due to interplanar interactions, shown in figure \ref{fig:Ldisp}.  The scattering was weak, and slices with relatively large widths,  $\pm 0.1$ in $h$ and $k$, were combined at symmetry-equivalent $\{hk\}$ for adequate statistics.  There is no clear dispersion along $l$ within the resolution of the binning, showing CoPS$_3$ to be a good approximation of a two-dimensional magnet.  The interplanar exchange, $J^\prime$, was thus set to be zero for the rest of the analysis.  A dedicated experiment with better statistics and resolution will be required to determine the size of $J^\prime$.

% Figure: MAPS slices along L at Gamma point
% end Figure: TAS Constant E = 2meV

\begin{figure}
  \includegraphics[width=3.5in]{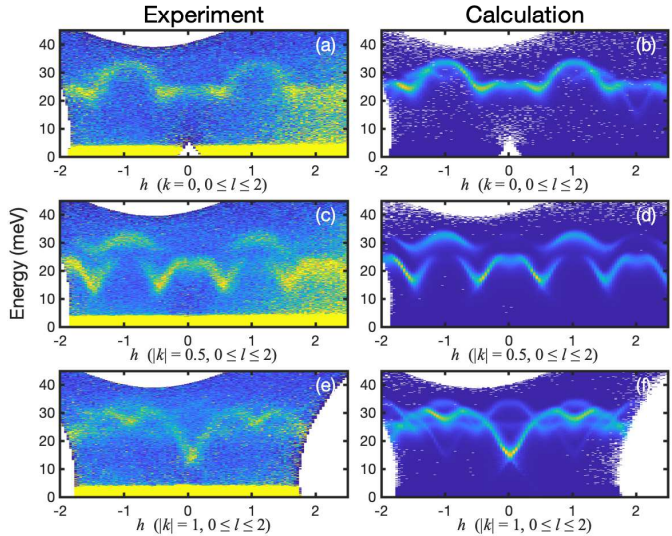}
  \caption{\label{fig:MAPSastar} Neutron inelastic scattering data from MAPS for slices parallel to the $a^{\star}$ axis.  Data and models have been integrated for $0 \le l \le 2$.  Slices centred at $\pm k$ were shown to be equivalent and have been combined for statistics with a slice width of  $\left|k\right| \pm 0.1$.  Slices along (a) $h0$, (c) $h\left|\frac{1}{2}\right|$ and (e) $h\left|1\right|$ are shown with corresponding calculations in (b), (d) and (f).}	
\end{figure}

\begin{figure}
  \includegraphics[width=3.5in]{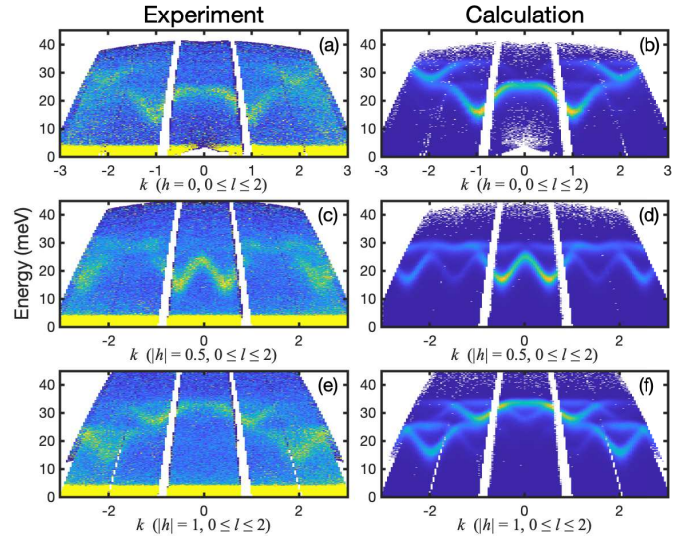}
  \caption{\label{fig:MAPSbstar} Neutron inelastic scattering data from MAPS for slices parallel to the $b^{\star}$ axis.  Data and models have been integrated for $0 \le l \le 2$.  Slices centred at $\pm h$ were shown to be equivalent and have been combined for statistics with a slice width of  $\left|h\right| \pm 0.1$.  Slices along (a) $0k$, (c) $\left|\frac{1}{2}\right|k$ and (e) $\left|1\right|k$ are shown with corresponding calculations in (b), (d) and (f).}	
\end{figure}

Data could thus be integrated along $l$ to improve statistics.  Symmetry-equivalent slices were also combined to improve statistics further.  Figure \ref{fig:MAPSastar} shows slices parallel to the ${a^{\star}}$ axis, integrated for $0 \le l \le 2$ and centred at different $k$.  Figure \ref{fig:MAPSbstar} shows similarly-binned slices parallel to the ${b^{\star}}$ axis centred at different $h$.   The dispersive modes are clearly visible, and slice trajectories may be mapped onto the $\left(a^{\star},b^{\star}\right)$ plane shown in figure \ref{fig:MSBZ}(d).

% Figure: MAPS slices along a*
% end Figure: MAPS slices along a*

% Figure: MAPS slices along b*
% end Figure: MAPS slices along b*

Although the 210 and $\frac{5}{2}10$ positions measured using three-axis spectrometers were inaccessible using the MAPS configuration, the slices contain numerous other symmetry-equivalent points.   

Data at integer values of $h$ in figures \ref{fig:MAPSastar}(a) and (c) correspond to $\Gamma$ points, and likewise at integer $k$ in figures \ref{fig:MAPSbstar}(a) and (c).  Modes at $\sim 15$ meV and $\sim 23$ are again observed, for example at $\left(h, k\right) = \left(1, 2\right)$ which is also a strong magnetic Bragg peak position.  A further two higher-energy modes are observed at positions like $\left(h, k\right) = \left(0,2\right)$ and $\left(1, 1\right)$.  These modes are strongest at silent magnetic Bragg peak positions.  No peak is observed at $18.5$ meV, supporting the conclusion that this is a spurious feature in the IN8 data.  Thus, a total of four spin-wave modes are present at the Brillouin zone centre, consistent with the conclusions from the analysis of neutron inelastic scattering from powdered samples \cite{CKim}.   

Data at half-integer $h$ in figures \ref{fig:MAPSastar}(a) and (c), and at integer $k$ in figure \ref{fig:MAPSbstar}(c), correspond to Y-points in the Brillouin zone and may be compared with figures \ref{fig:TAS}(b) and (h).  The MAPS data confirm the presence of the two modes observed on IN8, and show that these are the only two modes at this position.  The observations are again consistent with the previous analysis \cite{CKim}.

% ***************************
% ****** Analysis ******
% ***************************
\section{Analysis with linear spin wave theory \label{Sec:Ana}}
The experimental results showed well-defined and dispersive excitations at relatively large energies, consistent with a spin-wave picture.  The neutron scattering could thus be fitted and modelled using linear spin wave theory.  The dynamic structure factor, $S\left({\bf{Q},\omega}\right)$, was derived from a Hamiltonian using isotropic exchange parameters, $J_{ij}$ between moments $i$ and $j$, and two single ion anisotropies, $D^x$ and $D^z$:
\begin{equation}
  \begin{split}
  \hat{\mathcal{H}}=\frac{1}{2}\sum_{\left< ij \right>}&J_{ij} \left(S^x_iS^x_j + S^y_iS^y_j + S^z_iS^z_j \right) \\ 
  &+ D^x\sum_i\left[\left(S^{x}_i\right)^2-\left(S^{y}_i\right)^2\right] \\
  &+ D^z\sum_i\left(S^{z}_i\right)^2,
  \label{eq:Hamiltonian_Wildes}
  \end{split}
\end{equation}
where the \emph{x} axis is collinear with the aligned moment axis, the \emph{y} axis is parallel to the crystallographic $b$ axis, and \emph{z} is perpendicular to both \emph{x} and \emph{y}.  The single ion anisotropy is thus biaxial, with negative/positive $D^{x,z}$ defining an easy-axis/plane respectively.

CoPS$_3$ has four magnetic atoms per monoclinic unit cell.  The Hamiltonian is thus expressed as an $8 \times 8$ matrix which is not easily diagonalised analytically for an arbitrary reduced reciprocal lattice vector.  The SpinW software package for MATLAB\textsuperscript{\textregistered} \cite{SpinW} can provide analytical expressions for the spin wave energies when it is used with the MATLAB\textsuperscript{\textregistered} symbolic toolbox, however it was not able to provide general expressions in a reasonable time.  It was, though, able to provide analytical expressions from equation \ref{eq:Hamiltonian_Wildes} at specific high-symmetry points in the Brillouin zone.  The expressions are listed in Table \ref{tab:Eigenvalues}. 

%%%% Table with spin wave energies
\begin{table}
\caption{The analytic expressions for the eigenvalues at specific points in the Brillouin zone, as given by SpinW.  The label for each high-symmetry points, as defined in figure \ref{fig:MSBZ}(c) is listed beside its reduced lattice position.  The two or four modes at each point are numbered.  The parameters $D^x$ and $D^z$ are taken to be always negative, for uniaxial anisotropy, and positive, for planar anisotropy, respectively.  The parameters $U$ and $V$ for the eigenvalues at the Z point are defined in equations \ref{eq:U} and \ref{eq:V}}
\label{tab:Eigenvalues}
\begin{ruledtabular}
\begin{tabular}{ll}
$\Gamma$ & 0 0 0\\
1: & $2S\left(2D^x\left(D^x-D^z-J_1-4J_2-3J_3\right)\right)^{\frac{1}{2}}$ \\
2: & $2S\left(\left(D^x-D^z\right)\left(2D^x-J_1-4J_2-3J_3\right)\right)^{\frac{1}{2}}$ \\ 
3: & $2S\left(\left(D^x-D^z+2J_1-4J_2+2J^\prime\right)\left(2D^x+J_1-3J_3+2J^\prime\right)\right)^{\frac{1}{2}}$ \\
4: & $2S\left(2\left(D^x+J_1-2J_2+J^\prime\right)\left(D^x-D^z+J_1-3J_3+2J^\prime\right)\right)^{\frac{1}{2}}$ \\
\hline
Y & $\frac{1}{2} 0 0$ \\
1: & $2S\left(\left(2D^x-J_3+J^\prime\right)\left(D^x-D^z+J_1-2J_3+J^\prime\right)\right)^{\frac{1}{2}}$ \\
2: & $2S\left(\left(D^x-D^z-J_3+J^\prime\right)\left(2D^x+J_1-2J_3+J^\prime\right)\right)^{\frac{1}{2}}$ \\
\hline
Z & $0 \frac{1}{2} 0$ \\
1: & $S\left(2U - 2\left(V\right)^{\frac{1}{2}}\right)^{\frac{1}{2}}$ \\
2: & $S\left(2U + 2\left(V\right)^{\frac{1}{2}}\right)^{\frac{1}{2}}$ \\\\
\hline
C & $\frac{1}{2} \frac{1}{2} 0$ \\
1: & $2S\left(\left(2D^x+J^\prime\right)\left(D^x-D^z+J_1-3J_3+J^\prime\right)\right)^{\frac{1}{2}}$ \\
2: & $2S\left(\left(D^x-D^z+J^\prime\right)\left(2D^x+J_1-3J_3+J^\prime\right)\right)^{\frac{1}{2}}$ \\
\hline
B & $0 0 {\frac{1}{2}}$ \\
1: & $2S\left(\left(D^x-D^z+4J^\prime\right)\left(D^x-J_1-4J_2-3J_3+4J^\prime\right)\right)^{\frac{1}{2}}$ \\
2: & $2S\left(\left(D^x+4J^\prime\right)\left(D^z-D^x-J_1-4J_2-3J_3+4J^\prime\right)\right)^{\frac{1}{2}}$\\
3: & $2S\left(\left(D^x-D^z+2J_1-4J_2+2J^\prime\right)\left(D^x+J_1-3J_3+2J^\prime\right)\right)^{\frac{1}{2}}$\\
4: & $2S\left(\left(D^x+2J_1-4J_2+2J^\prime\right)\left(D^x-D^z+J_1-3J_3+2J^\prime\right)\right)^{\frac{1}{2}}$\\
  \end{tabular}
\end{ruledtabular}
\end{table}
% End table

The $xyz$ axes in equation \ref{eq:Hamiltonian_Wildes} are defined with respect to the spin direction and, with the exception of $y$, are not linked to the crystallographic axes.  The equations in table \ref{tab:Eigenvalues} are unchanged for any moment direction in the $\left(a,c\right)$ plane.  The magnetic structure for CoPS$_3$ has a small out-of-plane tilt of $\approx 9.5^{\circ}$ with respect to the crystallographic $a$-axis, but this was neglected in the SpinW data analysis and the moments were assumed to lie along $a$ such that $x$ and $a$ are collinear, as are the $z$ and $c^{\star}$ axes.  The assumption made no difference when calculating the spin wave energies, but impacts calculations for the neutron intensity which is given by those components of the moments that are perpendicular to the scattering vector $\bf{Q}$.  The impact, however, was small enough to be negligible.

Energy cuts at the high-symmetry points were extracted from the experimental data.  The data showed clear peaks which were fitted with Gaussians to determine the spin wave energies.  The resulting energies, with their corresponding $\bf{Q}$, are listed in table \ref{tab:DataFit}.
 
 % Table 2: Spin wave fits
\begin{table}
\caption{The data points, eigenvalues, and fitted parameters.  The eigenvalue numbering is defined with respect to the order in table \ref{tab:Eigenvalues}.  All energies are in meV.}
\label{tab:DataFit}
\begin{ruledtabular}
\begin{tabular}{ccccc}
{\bf{Q}} & k-vector & Eigenvalue & Measured energy & Fitted energy \\
\hline

 0 1  						& $\Gamma$ 	& 1 	& 14.5(1)	& 14.28 \\
1 2    					& $\Gamma$ 	& 1 	& 14.9(3) 	& " \\

0 0   						& $\Gamma$ 	& 2 	& 23.0(2) 	& 24.13 \\
2 0   						& $\Gamma$ 	& 2 	& 21.4(3) 	& " \\
1 2   						& $\Gamma$ 	& 2 	& 23.6(3) 	& " \\

1 0   						& $\Gamma$ 	& 3 	& 31.8(2) 	& 32.56 \\
0 1   						& $\Gamma$ 	& 3 	& 32.3(5) 	& " \\

0 2   						& $\Gamma$ 	& 4 	& 26.4(2) 	& 26.75 \\
1 1   						& $\Gamma$ 	& 4 	& 26.6(2) 	& " \\

$\frac{1}{2} 0$				& Y			& 1	& 23.3(1)	& 24.01 \\
$\frac{3}{2} 1$				& Y			& 1	& 23.4(2)	& " \\

$\frac{3}{2} 1$				& Y			& 2	& 29.6(2)	& 28.01 \\

$0 \frac{1}{2}$				& Z			& 1	& 22.0(1)	& 21.89 \\
$0 \frac{3}{2}$				& Z			& 1	& 22.1(3)	& " \\
$1 \frac{3}{2}$				& Z			& 1	& 22.2(1)	& " \\
$1 \frac{5}{2}$				& Z			& 1	& 21.9(1)	& " \\

$0 \frac{3}{2}$				& Z			& 2	& 30.8(3)	& 30.90 \\
$1 \frac{3}{2}$				& Z			& 2	& 31.0(2)	& " \\
$1 \frac{1}{2}$				& Z			& 2	& 31.2(1)	& " \\

$\frac{1}{2} \frac{1}{2}$		& C			& 1	& 15.1(1)	& 15.39 \\
$\frac{3}{2} \frac{1}{2}$		& C			& 1	& 15.5(1)	& " \\

$\frac{3}{2} \frac{1}{2}$		& C			& 2	& 27.1(2)	& 27.0 \\
$\frac{3}{2} \frac{3}{2}$		& C			& 2	& 28.5(2)	& " \\

\end{tabular}
\end{ruledtabular}
\end{table}
% End Table 2

The energies at each high symmetry point were attributed to appropriate eigenvalue equations, and then the ensemble was fitted with common parameters for the exchanges and anisotropies.    The attribution of the eigenvalue equations may seem arbitrary at first glance, however different attribution combinations quickly showed that only one gave results consistent with the data.  The eigenvalue attribution and fitted energies are also shown in table \ref{tab:DataFit}.

The best-fit exchange and anisotropy parameters are listed in table \ref{tab:Exchanges}.  Exchange interactions up to the third neighbour in the \emph{ab} plane, shown in figure \ref{fig:MSBZ}(a), were required to give an adequate fit which is consistent with similar conclusions for MnPS$_3$ \cite{Wildes98}, FePS$_3$ \cite{Lancon} and NiPS$_3$ \cite{Wildes22}.  Also consistently with the other members of the family, the second neighbour exchange, $J_2$, is very small, while $J_3$ is quite large and antiferromagnetic.  The first neighbour exchange, $J_1$, is ferromagnetic, as is the case in FePS$_3$ and NiPS$_3$, and it is the large $J_3$ that stabilises the antiferromagnetic zig-zag order.  The anisotropy energies, $D^{x,z}$, are large.  $D^z$ is positive, defining the $\left(a,b\right)$ plane to be easy, while the negative $D^x$ defines an easy direction in the plane parallel to the ordered collinear moments.  This differs to MnPS$_3$ and FePS$_3$ which can be modelled using only an easy-axis single-ion anisotropy.  It is similar to NiPS$_3$ which has the same magnetic structure with almost the same moment orientations, albeit having a much smaller $D^x/D^z$ ratio \cite{Wildes22}.

% Table 3: Fitted exchange parameters
\begin{table}
\caption{The fitted exchange parameters from the current data, and from previous analysis of neutron spectroscopy data from a powdered sample \cite{CKim}.  Negative $J$ denotes a ferromagnetic exchange.  $D^x$ and $D^z$ are the magnitudes of the single ion anisotropies along the ${a}$ and ${c^\star}$ axes respectively.  A negative $D$ denotes a uniaxial anisotropy, and a positive value denotes a planar anisotropy.  All exchange parameters and anisotropies are given in meV.}
\label{tab:Exchanges}
\begin{ruledtabular}
\begin{tabular}{cccc}
%$S$		& $1.5$ \\
%$J_1$ 	& $-1.34\left(8\right)$\\
%$J_2$ 	& $0.10\left(5\right)$ \\
%$J_3$	& $3.0\left(1\right)$\\
%$\alpha$	& 1	 \\
%$J^\prime$& $0$ \\
%$D^x$	& $-1.54\left(6\right)$\\
%$D^z$ 	& $5.2\left(4\right)$ \\
& Current work  & Reference \cite{CKim}  \\
 \hline
$S$			& $1.5$				& $1.5$ \\
$J_1$ 		& $-1.37\left(7\right)$	& $-2.04$	\\
$J_2$ 		& $0.09\left(5\right)$		& $-0.26$	 \\
$J_3$		& $3.0\left(1\right)$		& $4.21$	\\
$\alpha$		& $-$				& 0.6			 \\
$J^\prime$	& 0					& 0 \\
$D^x$		& $-0.77\left(3\right)$	& $-2.06$	\\
$D^z$ 		& $6.07\left(4\right)$		& $-$		 \\
 \end{tabular}
\end{ruledtabular}
\end{table}
% End Table 3

SpinW is able to numerically diagonalise the Hamiltonian for arbitrary $\bf{Q}$, subsequently calculating the directional components of the magnetic fluctuations and the expected neutron scattering intensity.  The calculated scattering for the IN8 and IN20 measurements is shown in figures \ref{fig:TAS}(i) and (j).  Instrument resolution has not been included in the calculations, whose widths have been broadened by an arbitrary $\Delta E = 1$ meV.  The $S^{\perp}S^{\perp}$ calculations represent those fluctuations perpendicular to $\bf{Q}$ which compare directly and favourably with the unpolarized IN8 data in figures \ref{fig:TAS}(c) and (d).  The directional components for the spin fluctuations in and normal to the scattering plane are shown in figures \ref{fig:TAS}(i) and (j) as $M_Y$ and $M_Z$ respectively, agreeing broadly with the IN20 data in figures \ref{fig:TAS}(g) and (h).

SpinW is also compatible with HORACE \cite{Horace} and the combined software can be used to calculate the expected neutron scattering, including a convolution with the instrumental resolution.  Once the spin wave energies in table \ref{tab:DataFit} were fitted and the best estimates for $J$ and $D^{x,z}$ were determined, the combined software was used to calculate the expected neutron scattering for a series of slices through the experimental data using an isotropic $g = 2$.  The calculations are remarkably similar to the experimental data, and are shown beside the corresponding slices in figures \ref{fig:MAPSastar} and \ref{fig:MAPSbstar}.

% ********************************
% ****** Discussion ************
% ********************************
\section{Discussion}
The analysis presented in section \ref{Sec:Ana} differs slightly from the previous analysis of the neutron data from powdered CoPS$_3$ \cite{CKim}, however the conclusions are coherent.  The best fits to the powder data used a Hamiltonian with a combination of a spin-direction-dependent anisotropic exchange with a single-ion term:
\begin{equation}
\begin{split}
  \hat{\mathcal{H}}=\frac{1}{2}\sum_{\left< ij \right>}&J_{ij} \left(S^x_iS^x_j + S^y_iS^y_j + \alpha S^z_iS^z_j \right) \\
  &+ D^x\sum_i\left(S^{x}_i\right)^2,
  \label{eq:Hamiltonian_Kim}
\end{split}
\end{equation}
where the definitions for the axes $xyz$ are the same as for equation \ref{eq:Hamiltonian_Wildes}.  An easy-plane anisotropy is obtained by the exchange anisotropy parameter $\alpha < 1$ in equation \ref{eq:Hamiltonian_Kim} or $D^z > 0$ in equation \ref{eq:Hamiltonian_Wildes}.  A negative $D^x$ creates a uniaxial anisotropy along the aligned moment direction in both equations \ref{eq:Hamiltonian_Wildes} and \ref{eq:Hamiltonian_Kim}, but the magnitudes will differ by a factor 2 as, for symmetry reasons, the relevant single ion term in equation \ref{eq:Hamiltonian_Wildes} includes a hard axis along the $y$ direction.   

\begin{figure}
  \includegraphics[width=3.5in]{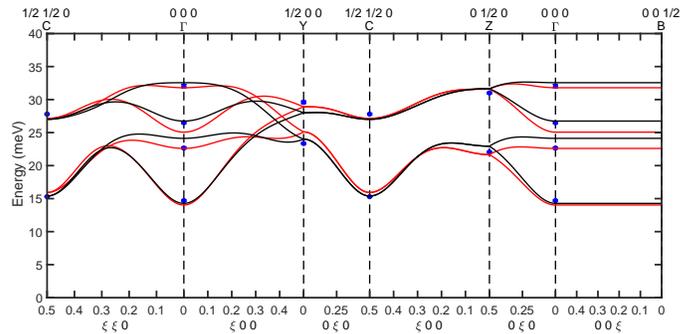}
  \caption{\label{fig:Spag} The spin wave dispersions calculated with the fitted exchange parameters for CoPS$_3$.  The red lines shows the dispersion using the parameters from Kim \emph{et al.} \cite{CKim}, derived from fitting neutron data from powders using equation \ref{eq:Hamiltonian_Kim}, while the black lines show the results from the current work using equation \ref{eq:Hamiltonian_Wildes}.  The data points at the high symmetry positions correspond to an average of the fitted spin wave energies for each of the equivalent eigenvalues listed in table \ref{tab:DataFit}.}	
\end{figure}

Phenomenologically, the choice of Hamiltonian is somewhat arbitrary as, for appropriate parameters, equations \ref{eq:Hamiltonian_Wildes} and \ref{eq:Hamiltonian_Kim} can be used to fit the data with equivalent quality.  For comparison, the parameters determined from analysing the neutron powder data are included in table \ref{tab:Exchanges} \cite{CKim}. The calculated spin wave dispersions for both sets of parameters are shown in figure \ref{fig:Spag} along with data points derived from appropriate averages of the fitted energies in table \ref{tab:DataFit}.  The two sets of parameters give very similar results, thus the experimental results and analysis on single crystals of CoPS$_3$ confirm the conclusions from the previous work on powdered samples.  Equation \ref{eq:Hamiltonian_Wildes} has an advantage that SpinW could give compact analytic expressions for the spin wave energies, shown in Table \ref{tab:Eigenvalues}, in a reasonable time, which was not the case for equation \ref{eq:Hamiltonian_Kim}.  Further motivation for using equation \ref{eq:Hamiltonian_Wildes} to model the data comes from considering the electronic ground state of the Co$^{2+}$ ions. 

% Figure: Spaghetti plot
% end Figure: Spaghetti plot

The most important point regarding the spin dynamics in CoPS$_3$ concerns the absence of a Kramers-doublet ground state arising from an effective $\mathcal{J}_\textrm{eff} = 1/2$ angular momentum state that might be expected from a $d^7$ ion in an octahedral environment \cite{Liu, Sano}.  A pertinent discussion has been presented by Winter \cite{Winter} involving the influence of the crystal electric field and spin-orbit coupling.  The crystal field of the octahedral environment of the S$^{2-}$ ions in CoPS$_3$ splits the $3$d orbitals of Co$^{2+}$ into a low-energy three-fold degenerate $t_{2g}$ orbital and a high-energy double-degenerate $e_g$ orbital with zero orbital momentum, split by a large energy expected to be of order $\sim 1$ eV.  The octahedra in CoPS$_3$ are distorted, having a trigonal elongation along the $c^{\star}$ axis \cite{Ouvrard85, Wildes17, CKim}, which further splits the $t_{2g}$ orbital into a singlet $a_{1g}$ with zero orbital momentum and, at a lower energy proportional to a term $\Delta_2$, a doublet $e^{\pi}_g$.  The $3d^7$ electrons of the Co$^{2+}$ fully fill the $e^{\pi}_g$ doublet and leave one unpaired electron in the $a_{1g}$ singlet and one in each $e_g$ orbital, giving a high-spin $S = 3/2$ with a quenched orbital moment.  The spin-orbit coupling must thus be considered in second order.  Furthermore, if the strength of the spin-orbit coupling, $\lambda$, is small compared to $\Delta_2$, i.e. $\Delta_2/\lambda \ge 1$, the $\mathcal{J}_\textrm{eff} = 1/2$ and $\mathcal{J}_\textrm{eff} = 3/2$ states merge and the net effect can be considered as a residual easy-plane single-ion anisotropy.  The picture is supported by DFT calculations and orbital-filling considerations \cite{Koo} which correctly reproduced the `zig-zag' {\bf{k}}$_\textrm{M} = \left[010\right]$ antiferromagnetic structure for CoPS$_3$ with the preferred moment direction perpendicular to the {\bf{c$^{\star}$}} axis.

Equation \ref{eq:Hamiltonian_Wildes} with isotropic exchange and a dominant easy-plane anisotropy provides hence a more natural description in view of the electronic ground state of Co$^{2+}$ in CoPS$_3$.  The anisotropic exchange parameters in equation \ref{eq:Hamiltonian_Kim} are appropriate for an effective $\mathcal{J}_\textrm{eff} = 1/2$ angular momentum state, which would be expected for an undistorted or a trigonally compressed octahedron and which could result in much sought-after Kitaev coupling.  CoPS$_3$ does not have this state, rather having an orbital singlet ground state, and hence the spin-orbit coupling results in a second order perturbation to a single-ion anisotropy and an anisotropic $g$-tensor.

The conclusion that the Co$^{2+}$ forms an orbital singlet due to the distortion is further reinforced as the Co$^{2+}$ ions have a monoclinic symmetry, lower than the trigonally-distorted octahedra, for which all orbital states are singlets.  This lower symmetry is important to explain the anisotropy, as trigonally-distorted octahedra have a planar symmetry about the quantisation axis and will not give rise to the biaxial anisotropy required to fit the CoPS$_3$ spectra.  

Estimates for Stevens parameters have been previously calculated for CoPS$_3$ with the appropriate local symmetry for the Co$^{2+}$ ions using first-order perturbation for spin $S = 3/2$ \cite{Wildes17}.  The calculations were based on the measured asymmetry in the paramagnetic susceptibility, giving values of $A^0_2 = -10.9$ K and $A^2_2\left(c\right) = -32.7$ K.  A value for $A^2_2\left(s\right)$ could not be explicitly determined but was assumed to be small.  The crystal field levels have not been comprehensively measured in CoPS$_3$, although neutron spectroscopy data suggest a feature at 70 meV \cite{CKim_21b}.  Future experiments are necessary to fully map the crystal fields.

As previously mentioned, the spin-orbit coupling in second-order perturbation leads to an anisotropy in the $g$ factors \cite{Winter}.  Previous efforts to quantify $g$ have focused on electron spin resonance (ESR) measurements of Co$^{2+}$-substituted in CdPS$_3$ \cite{Long, Hefni}.  The analysis was based on an effective spin $S_\textrm{eff} = 1/2$ and found $g_{\|} ~\sim 5$ and $g_\perp \sim 4$ for fields parallel to and perpendicular to $\bf{c^\star}$ respectively \cite{Zheng}.  Reanalysing the data with $S = 3/2$ has no effect on $g_{\|}$, but halves $g_\perp$ to become $\sim 2$ \cite{Long}.  These values are similar to the predictions from theory, giving $g_{\|} \approx 4$ and $g_\perp \approx 2$ in the limit that $\Delta_2/\lambda$ is large \cite{Winter}.  Further evidence for $g_{\|} > 2$ comes from the sizes of effective moment determined from the paramagnetic susceptibility and from the ordered moment determined by neutron diffraction \cite{Wildes17} which are slightly larger than expected for spin-only and $g = 2$.

The value for $g$ does not affect the spin wave energies, but does affect the calculated neutron scattering intensity.  The calculations in figures \ref{fig:TAS}, \ref{fig:MAPSastar} and \ref{fig:MAPSbstar} used an isotropic $g=2$ and $S = 3/2$.  The agreement between calculation and data is satisfactory, which implies that $g$ in CoPS$_3$ may be more isotropic than expected from the theory and the ESR results.  The two conclusions are not necessarily contradictory.  The theory and the ESR results consider local moments, which is appropriate for dilute substitution in a CdPS$_3$ matrix as the Co$^{2+}$ ions may be treated as isolated impurities.  CoPS$_3$ is much more concentrated and the influence of the molecular field on the electronic states must also be considered, especially as the exchange parameters in table \ref{tab:Exchanges} are relatively large and must be considered up to the third neighbour in-plane.  The molecular field term may cause the $g$ factor to become more isotropic.   A full treatment of CoPS$_3$, along with other members of the TM-PS$_3$ family, will be considered in future work.

% ***************************
% ****** Conclusions ******
% ***************************
\section{Conclusions}
Neutron inelastic scattering has been used to measure the spin wave spectrum of CoPS$_3$.  Four clear and dispersive spin wave branches were observed, and the data could be fitted with a model derived from a Hamiltonian with isotropic exchange interactions and a biaxial single-ion anisotropy.  The data are consistent with the results and conclusions of previous neutron spectroscopy data from powdered samples.  Exchange interactions up to the third in-plane neighbour needed to be included in the fitting, with resulting values of $-1.37$ meV, $0.09$ meV and $3.0$ meV for first, second and third neighbours respectively.  The anisotropy was defined by an easy-plane term and an easy-axis term within the plane with values of $6.07$ meV and $-0.77$ meV respectively.  The results are consistent with high-spin $S = 3/2$ spin waves, which may be understood based on the effects of the low point symmetry of the Co$^{2+}$ sites on the crystal electric field levels and the strength and range of the exchange.

%Acknowledgements
\begin{acknowledgments}
We thank the Institut Laue-Langevin, experiment numbers TEST-3025 \cite{Wildes_IN3_July19}, 4-01-1632 \cite{Wildes_IN8_July19} and 4-01-1651 \cite{Wildes_IN20_Sep20}  for the allocation of neutron beam time.  Experiments at the ISIS Neutron and Muon Source were supported by a beamtime allocation RB2010386 \cite{Wildes_MAPS_Oct20} from the Science and Technology Facilities Council.  We also thank Dr. A. Piovano and Dr. A. Ivanov, F. Charpenay and V. Gaignon for assistance with the IN8 experiment. and Dr. R. Ewings and Dr. M. D. Le for their assistance with SpinW and HORACE.  ARW sincerely thanks Dr. C. Stock, Dr. M. Zhitormirsky and Dr. R. Ballou for enlightening discussions.
\end{acknowledgments}

% Appendix
\appendix
\section{\label{app:evals} Expressions for $U$ and $V$}
The expressions for the variables $U$ and $V$, used to calculate the spin wave energies at the Z points in Table \ref{tab:Eigenvalues}, are:
\begin{equation}
 \label{eq:U}
  \begin{split}
 U =& \left(D^x-D^z\right)\left(2D^x+J_1-4J_2-3J_3+2J^{\prime}\right) \\
       & +2D^x\left(D^x-D^z+J_1-4J_2-3J_3+2J^{\prime}\right) \\
       & + 2J_1^2 + 8J_2^2 + 4J_3^2 + 4J^{\prime 2} \\
       & - 4J_1J_2 - 4J_1J_3 + 6J_1J^{\prime} \\
       & +12J_2J_3  - 8J_2J^{\prime} - 6J_3J^{\prime}
  \end{split}
\end{equation}
and
\begin{equation}
 \label{eq:V}
  \begin{split}
 V =& 2D^x\left(D^x-D^z\right)\left(4J_1 + 4J^{\prime}\right)^2  \\
       &+ \left(D^x-D^z\right)^2\left(\left(J_1 + J_3\right)^2 + 4\left(J_1 + J^{\prime}\right)^2\right) \\
       &+ 8\left(3D^x - Dz\right)\left(J_1 + J^{\prime}\right)^2\left(J_1 - 4J_2 - 3J_3 + 2J^{\prime}\right) \\
       &- 16( 2J_1^3J_2 + 2J_1^3J_3 - J_1^3J^{\prime} - 4J_1^2J_2^2 - 6J_1^2J_2J_3\\
       &  + 8J_1^2J_2J^{\prime} - 2J_1^2J_3^2 + 7J_1^2J_3J^{\prime} - 3J_1^2J^{\prime 2}\\
       &  - 8J_1J_2^2J^{\prime} - 12J_1J_2J_3J^{\prime} + 10J_1J_2J^{\prime 2} - 4J_1J_3^2J^{\prime}\\
       &  + 8J_1J_3J^{\prime 2} - 3J_1J^{\prime 3} - 4J_2^2J^{\prime 2} - 6J_2J_3J^{\prime 2} \\
       & + 4J_2J^{\prime 3} - 2J_3^2J^{\prime 2} + 3J_3J^{\prime 3} - J^{\prime 4})
   \end{split}
\end{equation}

\clearpage
%\bigskip

\bibliography{MPS3_arxiv}% Produces the bibliography via BibTeX.
%\bibliography{MPS3}% Produces the bibliography via BibTeX.

\end{document}